\documentclass[journal]{IEEEtran}
\usepackage{cite}
\usepackage{epsfig,latexsym,amssymb,amsmath,graphics,color,bbm}
\usepackage{amsfonts}
\usepackage{subfigure}
\usepackage{multicol}
\usepackage{graphicx,subfigure}
\usepackage[linesnumbered,ruled]{algorithm2e}
\usepackage{float}
\usepackage{stfloats}
\usepackage{epstopdf}
\newtheorem{prop}{Proposition}

\newtheorem{cor}{Corollary}

\newtheorem{lm}{Lemma}

\newtheorem{thm}{Theorem}

\newcommand{\be}{\begin{eqnarray}}
\newcommand{\ee}{\end{eqnarray}}
\newcommand{\benn}{\begin{eqnarray*}}
\newcommand{\eenn}{\end{eqnarray*}}
\def\IR{\rm I \kern-0.20em R}

\newcommand{\bthm}{\begin{thm}}
\newcommand{\ethm}{\end{thm}}

\newcommand{\bcor}{\begin{cor}}
\newcommand{\ecor}{\end{cor}}
\newcommand{\bprop}{\begin{prop}}
\newcommand{\eprop}{\end{prop}}
\newcommand{\blm}{\begin{lm}}
\newcommand{\elm}{\end{lm}}
\newcommand{\beq}{\begin{equation}}
\newcommand{\eeq}{\end{equation}}
\newcommand{\ber}{\begin{eqnarray}}
\newcommand{\eer}{\end{eqnarray}}

\newcommand{\bproof}{\begin{proof}}
\newcommand{\eproof}{\end{proof}}

\newcommand{\bit}{\begin{itemize}}
\newcommand{\eit}{\end{itemize}}
\newcommand{\ben}{\begin{enumerate}}
\newcommand{\een}{\end{enumerate}}
\newcommand{\bdesc}{\begin{description}}
\newcommand{\edesc}{\end{description}}
\newcommand{\beqarrn}{\begin{eqnarray*}}
\newcommand{\eeqarrn}{\end{eqnarray*}}
\newcommand{\bproofof}{\begin{proofof}}
\newcommand{\eproofof}{\end{proofof}}
\newenvironment{rem}{\begin{trivlist}\item[]{\bf
Remark:}\hspace{4mm}}{\end{trivlist}}
\newcommand{\brem}{\begin{rem}}
\newcommand{\erem}{\end{rem}}
\newenvironment{rems}{\begin{trivlist}\item[]{\bf
Remarks}\begin{itemize}}{\end{itemize}\end{trivlist}}
\newcommand{\brems}{\begin{rems}}
\newcommand{\erems}{\end{rems}}
\newtheorem{fact}{Fact}
\newcommand{\bfact}{\begin{fact}}
\newcommand{\efact}{\end{fact}}
\newtheorem{examp}{Example}
\newcommand{\bexamp}{\begin{examp}\rm}
\newcommand{\eexamp}{\end{examp}}
\newtheorem{defn}{Definition}
\newcommand{\bdefn}{\begin{defn}\rm}
\newcommand{\edefn}{\end{defn}}

\newtheorem{alg}{Algorithm}
\newcommand{\balg}{\begin{alg}}
\newcommand{\ealg}{\end{alg}}

\newtheorem{prob}{Problem}
\newcommand{\bprob}{\begin{prob}}
\newcommand{\eprob}{\end{prob}}

\newcommand{\bvtm}{\begin{verbatim}}
\newcommand{\bfig}{\begin{figure}}
\newcommand{\efig}{\end{figure}}
\newcommand{\bcen}{\begin{center}}
\newcommand{\ecen}{\end{center}}

\long\def\comment#1{}

\def \n2{{N_0 \over 2}}

\def \h5{\hspace{0.5in}}

\newcommand{\dff}{\stackrel{\triangle}{=}}

\def\IR{\mathbb R}

\makeatletter
\renewcommand\normalsize{%
	\@setfontsize\normalsize\@xpt\@xiipt
	\abovedisplayskip 2\p@ \@plus1\p@ \@minus2\p@
	\abovedisplayshortskip \z@ \@plus2\p@
	\belowdisplayshortskip 2\p@ \@plus1\p@ \@minus2\p@
	\belowdisplayskip\abovedisplayskip
	\let\@listi\@listI}
\makeatother
\renewcommand{\baselinestretch}{1.0}

\newtheorem{theorem}{Theorem}
\newtheorem{lemma}{Lemma}

\input{epsf}

	\title{ Clipping noise approximate analysis and power allocation for photon-detection-based DCO-OFDM and ACO-OFDM }
	
	\author{Zhimeng Jiang, Chen Gong, and Zhengyuan Xu
		\thanks{This work was supported by Key Program of National Natural Science Foundation of China (Grant No. 61631018) and Key Research Program of Frontier Sciences of CAS (Grant No. QYZDY-SSW-JSC003).}
		\thanks{The authors are with Key Laboratory of Wireless-Optical Communications, Chinese Academy of Sciences, University of Science and Technology of China, Hefei, Anhui 230027, China. 
			Email: zhimengj@mail.ustc.edu.cn, \{cgong821, xuzy\}@ustc.edu.cn.}}

\begin{document}
	\maketitle
	\begin{abstract}
		The clipping noise of the photon-level detector for both direct current-biased optical OFDM (DCO-OFDM) and asymmetrically clipped optical OFDM (ACO-OFDM) is investigated. Based on Bussgang theorem and central limit theorem (CLT), we obtain the approximate closed-form SNR of each subcarrier, based on which we further formulate the power allocation among the subcarriers. Numerical results show that the SNR obtained from theoretical analysis can well approximate that obtained from simulation results, and uniform power allocation suffices to perform close to the optimized power allocation from Genetic Algorithm (GA) with significantly reduced computational complexity.
	\end{abstract}

	\IEEEpeerreviewmaketitle
	\section{Introduction}
	Current optical wireless communication (OWC) serves as a feasible candidate for medium range data transmission where the radio-frequency (RF) radiation is prohibited \cite{randel2010advanced}. Two typical OFDM approaches are adopted, direct current-biased optical OFDM (DCO-OFDM) with a DC bias, and asymmetrically clipped optical OFDM (ACO-OFDM) with the negative component clipped \cite{armstrong2006power,armstrong2009ofdm,dardari2000theoretical}. Experimental comparison of different bit and power allocation algorithms for visible light communications (VLC) system using DC-biased optical OFDM is presented in \cite{bykhovsky2014experimental}. The power of worst-case residual clipping noise in LACO-OFDM is investigated in \cite{zhang2018worst} for VLC waveform signals. The time-domain signal is clipped from both sides, including downward and upward clipping caused by insufficient DC bias and physical limitation of transmitted optical power, especially for the eye safety \cite{en200862471}.
	
	On the other hand, photon-level detector, such as photomultiplier tube (PMT) and single photon avalanche diode (SPAD) \cite{almer2015spad}, can be applied in the scenario of weak light reception power, such as ultraviolet communication \cite{ding2009modeling} and visible light communication under extremely weak transmission signal and ambient light power. The clipping noise and signal shaping for OFDM is investigated in \cite{dimitrov2012clipping}, which shows that non-linear LED I-V characteristic can be compensated by pre-distortion and a linear characteristic can be obtained over a limited range. Poisson channel, couped with signal-dependent noise, is typical for photon-level receiver in optical wireless communication.  The photon-level signal characterization without top clipping for DCO-OFDM has been investigated in \cite{li2015optical}, but it is still not clear that the effect of clipping noise incorporating the signal-dependent noise on system performance. It would be necessary to characterize the received signal with clipping noise under different top clipping levels for DCO-OFDM and ACO-OFDM with a photon-level detector due to limited linear range of LED, and investigate the performance of DCO-OFDM and ACO-OFDM under photon-level detection. The photon-level signal characterization without top clipping for DCO-OFDM has been investigated in \cite{li2015optical}, The contribution of this work beyond \cite{li2015optical} lies in characterizing the received signals with top clipping and optimizing the power allocation among the subcarriers for both DCO-OFDM and ACO-OFDM.
	
	In this letter, we investigate the photon-level signal characterization with clipping for both DCO-OFDM and ACO-OFDM. We provide closed-form SNR for each subcarrier at the receiver and formulate an optimization problem to maximize the system total rate. The closed-form SNR is verified by the numerical results. Moreover, it is observed that uniform power allocation among the subcarriers can perform close to the optimized power allocation obtained by Genetic algorithm, with significantly reduced computational complexity.
	\section{System Model}\label{sec.SystemModel}
	\subsection{LED Transmitter}
	\begin{figure*}
		\centering
		{\includegraphics[width = 2 \columnwidth]{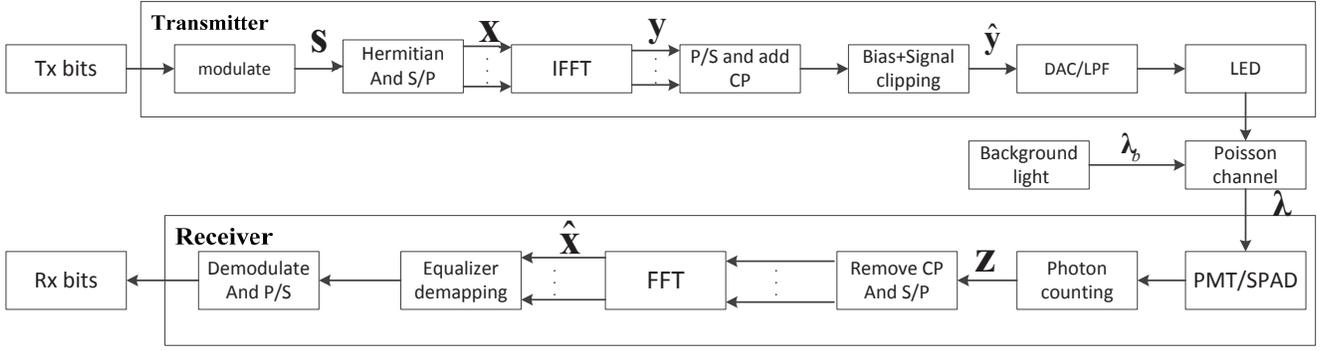}}
		\caption{Block diagram of DCO-OFDM system}
		\label{fig_OFDMsystem}
	\end{figure*}
	The DCO-OFDM system model and main notations are shown in Fig.\ref{fig_OFDMsystem}.
	Consider the transmission with DCO-OFDM and ACO-OFDM. The signals on each subcarrier, denoted as $x_{k}$, are given by $x_{k} = s_{k}w_{k}$ for $k = 0,1,\cdots,N-1$, where $w_{k}$ is the linear scale coefficient of the $k^{th}$ subcarrier and  $s_{k}$ is the symbol of $k^{th}$ subcarrier after modulation with $\mathbb{E}[s^2_k] = 1$. For DCO-OFDM, symbols $x_k$ for $k = 1,\cdots,N/2-1$, are mapped to subcarrier $k$; and for ACO-OFDM symbols, $x_k$ for $k = 1,3,\cdots,N/2-1$ are mapped to subcarrier $k$, whereas the symbols on even subcarriers are set to be zero. ACO-OFDM is energy-saved at the cost of bandwidth compared with DCO-OFDM. Hermitian symmetry is adopted for the rest half subcarriers to guarantee real-valued symbols after the IFFT, given by
	$y_{n}=\sum_{k=0}^{N-1}x_{k}e^{\frac{j2\pi kn}{N}}$
	where $y_{n}$ is the time-domain symbol. For DCO-OFDM, a DC bias is added to make signal unipolar with LED maximum power $y_{max}$, given by
	\begin{eqnarray}
	B_{DC}=\epsilon_{B} \sigma_y, \quad y_{max}=\epsilon_{top} \sigma_y,
	\end{eqnarray}
	where $\epsilon_{B}$  and $\epsilon_{top}$ are defined as the bias level and top level and $\sigma_y\triangleq\sqrt{\mathbb{E}{[y_n^{2}}]}=\sqrt{\sum_{k=0}^{N-1}w_k^2}$. The signal after adding DC bias is given by $y_{n}^{bias}=y_{n}+B_{DC}$.
	
	For ACO-OFDM, only the positive parts are transmitted and can be recovered based on original odd symmetry signal. The definition of $\epsilon_{top}$ of ACO-OFDM is similar to that in DCO-OFDM. Thus the clipped signal is given by $\hat {y}_{n}=C(y_{n})=y_{n}\mathbbm{1}\{0\leq y_{n}\leq y_{max}\}+y_{max}\mathbbm{1}\{y_{n}> y_{max}\}$, where $\mathbbm{1}$ is a indicator function.
	
	\subsection{Channel Model}
	Assume low transmission power or large path loss such that continuous waveform cannot be detected and a photon-counting receiver needs to be deployed. The detected signal satisfies a Poisson distribution with mean $\lambda _{n}=\alpha y_{n}^{r}+\lambda _{b}$, where $y_{n}^{r}$ denotes received power, $\lambda_{b}$ denotes the mean number of background radiation and dark current, and $\alpha$ denotes the ratio of mean number of photons over the signal power. Note that we have $\alpha = \frac{\tau}{h\nu}$,
	where $\tau$ denotes symbol duration, and $h$ and $\nu$ denote the Planck's constant and the frequency of the optical signal, respectively. The number of detected photons, denoted as $z_{n},$ is characterized by probability $\mathbb{P}\Big(z_{n}=k_1\Big) = \frac{\lambda_{n}^{k_1}}{k_1!}e^{-\lambda_{n}}$ \cite{li2015optical,wyner1988capacity}.
	Due to the low-pass filtering characteristics of the LED, different OFDM subcarriers may have different link gains, denoted as $g_{k}$ for $k=0, 1, \cdots, N-1$, which incorporates LED low-pass filtering and the link gain between the transmitter and the receiver. Assume perfect knowledge on the subcarrier gains at the transmitter.

	\section{Clipping Noise Analysis and Power Allocation}\label{sec.ReceiverDesign}
	
	\subsection{Performance Analysis with Clipping Noise}
	Note that symbol $x_k$ can be estimated based on the FFT output of $z_n$, denoted as $\hat x_k$. According to Bussgang theorem, the clipping function $C(\cdot)$ can be expressed as $\hat{y}_n=K\cdot y_n^{bias}+n_c(n)$, where $n_c(n)$ is the time domain clipping noise, uncorrelated with $y_n^{bias}$, and $K=\frac{\mathbb{E}[\hat{y}_ny_n^{bias}]}{\mathbb{E}[(y_n^{bias})^2]}$ is the scaling factor. We adopt identically and independently distributed Gaussian clipping noise assumption for both DCO-OFDM and ACO-OFDM \cite{dimitrov2012clipping}. We have the following results on the noise power on each subcarrier.
	\begin{theorem}\label{the.clip}
		For DCO-OFDM, the variance of $\hat x_k$ on subcarrier $k$ is given by
		\be
		\mathbb{D}[\hat{x}_{k}]=\frac{1}{N}[\alpha g_0(KB_{DC}+\mu)+\lambda_b]+\frac{\alpha^2\sigma^2}{N}|g_k|^2,
		\ee
		where
		\begin{flalign}
		K&= \{\epsilon_B[\phi(\epsilon_B)-\phi(\epsilon_{top}-\epsilon_B)]+(1+\epsilon_B^2)Q(-\epsilon_B)\nonumber \\&\quad-(1+\epsilon_B^2-\epsilon_{top}\epsilon_B)Q(\epsilon_{top}-\epsilon_B)]\}/(1+\epsilon_B^2);\\ \label{eq.k}
		\phi (u)&=\frac{1}{\sqrt{2\pi }}\exp(-\frac{u^2}{2}), Q(u)=\int_{u}^{+\infty }\phi (t)dt; \nonumber\\
		\mu&=\mathbb{E}[n_c(n)]=\sigma_y[(1-K)\epsilon_B(1-\epsilon_B)Q(\epsilon_B)\nonumber \nonumber\\
		&\quad+(\epsilon_{top}-\epsilon_{B}-1)Q(\epsilon_{top}-\epsilon_{B})];\nonumber\\ 
		\sigma^2&=\sigma_y^2[\epsilon_B\phi (-\epsilon_B)-(\epsilon_B+\epsilon_{top})\phi(\epsilon_{top}-\epsilon_B)\nonumber\\ 
		&\quad+(1+\epsilon_B^2)Q(-\epsilon_B)+(\epsilon_{top}^2-\epsilon_B^2-1)Q(\epsilon_{top}-\epsilon_B)]\nonumber\\ 
		&\quad-K^2(\sigma_y^2+B_{DC}^2)-\mu^2; \nonumber
		\end{flalign}
		On the other hand, for ACO-OFDM, the variance of $\hat x_k$ on subcarrier $k$ is given by
		\be
		\mathbb{D}[\hat{x}_{k}]=\frac{1}{N}\Big(\alpha^2\sigma^2|g_k|^2+\alpha g_0(K\frac{\sigma_y}{\sqrt{2\pi}}+\mu)+\lambda_b\Big),
		\ee
		where $K=1-2Q(\epsilon_{top})$, $\sigma^2_y =\sum^{N-1}_{k=1}w^2_k$ and
		\begin{flalign}
		\mu&=\sigma_y[-\phi (\epsilon_{top})+\epsilon_{top}Q(\epsilon_{top})+\frac{1-K}{\sqrt{2\pi}}];\nonumber\\
		\sigma^2&=\sigma_y^{2}[\frac{1-K^2}{2}+(\epsilon_{top}^2-1)Q(\epsilon_{top})-\epsilon_{top}\phi(\epsilon_{top})]-\mu^2.\nonumber
		\end{flalign}
		\begin{proof}
			Please refer to Appendix \ref{app.1}.
		\end{proof}
	\end{theorem}	
	
	Note that $\frac{\hat{x}_k}{\alpha Kg_k}$ is an unbiased estimate of $x_k$ given $\alpha$, $K$ and $g_k$. Thus, define $SNR_k\dff\frac{|\mathbb{E}[x_k]|^2}{\mathbb{E}[|\frac{\hat{x}_k}{\alpha Kg_k}-x_k|^2]}$ to evaluate the quality of estimate $\hat{x}_k$. For DCO-OFDM, we have that
	\begin{flalign} 
	\mathbb{E}[y_n^r]=K[g_0B_{DC}+\sum_{k=1}^{N-1}g_k x_k e^{j\frac{2\pi nk}{N}}]+\mu g_0,
	\end{flalign} 
	and the expectation 
	\begin{flalign} 
	\mathbb{E}[z_n]&=\mathbb{E}[\alpha y_n^r+\lambda_b]\\
	&=\alpha g_0(KB_{DC}+\mu)+\alpha K\sum_{k=0}^{N-1}g_k x_k e^{j\frac{2\pi nk}{N}}+\lambda_b.\nonumber
	\end{flalign} 
	Furthermore, via taking FFT on $z_n$, we have
	\be 
	\mathbb{E}[\hat{x}_{k}]=\frac{1}{N}\sum_{n=0}^{N-1}\mathbb{E}[z_n]e^{-j\frac{2\pi nk}{N}}=\alpha K g_k x_k, \text{for}\quad k\neq0.
	\ee 
	Note that estimate $\hat{x}_k$ is a unbiased estimate of $x_k$ given $\alpha$, $K$, dependent on transmitter, and $g_k$, thus, the SNR of subcarrier $k$ for DCO-OFDM is given by,
	\be\label{eq.DCOSNR}
	SNR_k^{DCO}&=&\frac{|\mathbb{E}[\hat{x}_k]|^2}{\mathbb{D}[\hat{x}_k]}\nonumber\\&=&\frac{N\alpha ^2 K^2 w_k^2 |g_k|^2}{\alpha^2\sigma^2|g_k|^2+\alpha g_0(KB_{DC}+\mu)+\lambda_b}.
	\ee	
	For ACO-OFDM, we have
	\be 
	\mathbb{E}[y_n^r]=\frac{K}{2}\sum_{k=0}^{N-1}g_k x_k e^{j\frac{2\pi nk}{N}}+K\sum_{k=0}^{N-1}g_k D_k e^{j\frac{2\pi nk}{N}}+\mu g_0,
	\ee
    Note that $D_k$ is equivalent to $0$ for odd $k$, we have the expectation of $\hat{x}_k$
    \be 
    \mathbb{E}[\hat{x}_{k}]&=&\frac{1}{N}\sum_{n=0}^{N-1}\mathbb{E}[\alpha y_n^r+\lambda_b]e^{-j\frac{2\pi nk}{N}}\nonumber\\&=&\frac{\alpha K g_k x_k}{2} \quad\text{for}\quad k=\text{odd}.
    \ee 
	Similar, estimate $\hat{x}_k$ is a unbiased estimate of $x_k$ given $\alpha$, $K$ and $g_k$. Thus the SNR of odd subcarrier $k$ for ACO-OFDM is given by,
	\be\label{eq.ACOSNR}
	SNR_k^{ACO}=\frac{N\alpha ^2 K^2 w_k^2 |g_k|^2}{4(\alpha^2\sigma^2|g_k|^2+\alpha g_0(K\frac{\sigma_y}{\sqrt{2\pi}}+\mu)+\lambda_b)}.
	\ee
	 
	Based on above analysis, the noise power consists of three parts, the clipping noise part, the Poisson noise part and the background radiation part, corresponding to the first item, second item and last item of denominator in Equations (\ref{eq.DCOSNR}) and (\ref{eq.ACOSNR}).

	\subsection{Power Allocation for Subcarriers}
	For transmission power constraint, the optical power is upper bounded by $P_{Tmax}$, i.e., $\mathbb{E}[\hat{y}_n]\leq P_{Tmax}$. For DCO-OFDM, the mean transmission power $\mathbb{E}[\hat{y}_n]$ is related to the bias $B_{DC}$ and the clipping, i.e., $\mathbb{E}[{\hat y}_n]=B_{DC}+\beta_{DCO}$, where $\beta_{DCO}$ is the optical power adjustment due to clipping, given by
	\begin{flalign}
	\beta_{DCO}=&\mathbb{E}[y_n^{clip}]=\sigma_y[\phi(\epsilon_B)-\phi(\epsilon_{top}-\epsilon_B)\nonumber \\
	&+(\epsilon_{top}-\epsilon_B)\phi(\epsilon_{top}-\epsilon_B)-\epsilon_B Q(\epsilon_B)].
	\end{flalign}
	On the other hand, for ACO-OFDM, we have $\mathbb{E}[{\hat y}_n]=\frac{\sigma_y}{\sqrt{2\pi }}+\beta_{ACO}$, where $\beta_{ACO}=\sigma_y[\epsilon_{top} Q(\epsilon_{top})-\phi(\epsilon_{top})]$.
	
	The system design aims to maximize the sum rate of each valid subcarrier $\log(1+SNR)$ due to approximate Gaussian noise in each subcarrier, subject to the transmission power constraint. It is justified by that what we concern is channel $\mathbf{x}\rightarrow\hat{\mathbf{x}}$ instead of $\hat{\mathbf{y}}\rightarrow\mathbf{z}$ and the
	frequency-domain signals on each subcarrier after taking FFT can be well approximated
	by Gaussian according to CLT while the received time-domain signals cannot be well approximated using Gaussian. Numerical results for 4-QAM modulation validate the approximate capacity formula $\log(1+SNR)$. For DCO-OFDM, it is formulated as follows,
	\begin{equation}
	\begin{aligned}
	& \underset{B_{DC},w_i}{\text{max}}
	& &\sum_{k=0}^{N/2-1}\log(1+SNR_{k}^{DCO}),\\
	& \text{s.t.}
	& &B_{DC}+\beta_{DCO}\leq P_{Tmax};\\
	&&&0<B_{DC}<y_{max}.
	\end{aligned}
	\end{equation}
	For ACO-OFDM, it is formulated as follows,
	\begin{equation}
	\begin{aligned}
	& \underset{w_i}{\text{max}}
	& &\sum_{k=1}^{N/4}\log(1+SNR_{2k-1}^{ACO}),\\
	& \text{s.t.}
	& &\frac{\sigma_y}{\sqrt{2\pi }}+\beta_{ACO}\leq P_{Tmax}.\\
	\end{aligned}
	\end{equation}
	
	\begin{lemma}\label{lemma.noncon}
		The constraint function for Problems (13) and (14) are non-convex.
		\begin{proof}
			Please refer to Appendix \ref{lemB}.
		\end{proof}	
	\end{lemma}
	
	According to lemma \ref{lemma.noncon}, Problems (13) and (14) are non-convex. For non-convex and continuous
	optimization problem with multiple variables, it cannot be solved via exhaustive search and we resort to standard
	genetic algorithm (GA) to solve it. It is seen that for both DCO-OFDM and ACO-OFDM, the SNR on subcarrier $k$ is linear with $w_k^2$. Given $\sigma^2_y = \sum^{N-1}_{k=1}w_k^2$, we observe that uniform $w_k^2$ can perform close to the optimized solution from GA.

	\section{Numerical Results}\label{sec.NumericalResults}	
	The linear scale $w_i$ and DC bias $B_{DC}$ are optimized for both DCO-OFDM and ACO-OFDM subject to the power constraint. The blue LED frequency response is obtained from experimental measurements, which shows the $3$dB bandwidth of $8.5$MHz by spectrum analyzer. Assume $64$ subcarriers for the OFDM. The subcarrier gains of the subcarriers $g_k$ incorporate the LED frequency gains and path gains, where those of the first $32$ subcarriers from the real experimental measurements, as shown in Table \ref{tab.subgain}, arranged in row by row from left to right. The gains of the rest $32$ subcarriers can be obtained based on Hermitian symmetry. The symbol rate is 20Mbps, and the mean number of background noise photons within each symbol duration $\lambda_b = 0.001$. Assume that $P_{Tmax}=0.1W$. The SNRs of 4-QAM DCO-OFDM and ACO-OFDM from both theoretical analysis (denoted as theo) and simulations (denoted as simu) with the linear scale $w_k=0.5$ for all  information-carried subcarriers are presented in Fig. \ref{fig_DCOSNR} and Fig. \ref{fig_ACOSNR}, respectively, for different values of $\epsilon_B$ and $\epsilon_{top}$. For DCO-OFDM, the SNR first increases and then decreases with the DC bias level $\epsilon_{B}$, as the Poisson noise component dominates for a large DC bias level. The gap between the theoretical predictions and the simulation results can reach more than $1$dB for small $\epsilon_B$ and $\epsilon_{top}$, which can be justified by the larger clipping noise with non-negligible correlation between the samples. For ACO-OFDM, the SNR increases with the top level $\epsilon_{top}$, and the performance gain becomes saturated when the top level raises above a threshold. For both DCO-OFDM and ACO-OFDM, the theoretical SNRs match well with the simulation results, which validates Gaussian approximation.
	\begin{figure}
		\setlength{\abovecaptionskip}{-0.3cm} 
		\centering
		{\includegraphics[width =0.9\columnwidth]{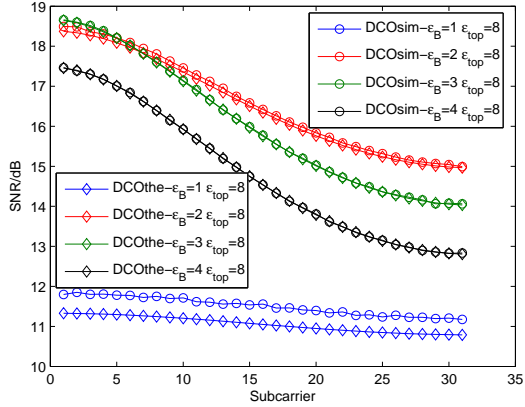}}
		\caption{The SNR results from both theoretical derivations and simulations for each subcarrier for 4-QAM DCO-OFDM.}
		\label{fig_DCOSNR}
	\end{figure}
	\begin{figure}
		\setlength{\abovecaptionskip}{-0.3cm} 
		\centering
		{\includegraphics[width =0.9\columnwidth]{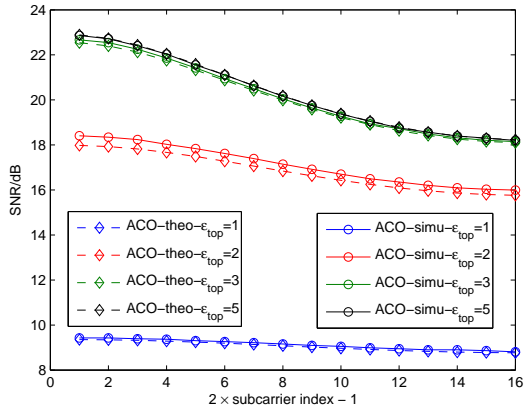}}
		\caption{The SNR results from both theoretical derivations and simulations for each subcarrier in 4-QAM ACO-OFDM.}
		\label{fig_ACOSNR}
	\end{figure}
	
	Moreover, it is shown that residual error $\hat{\mathbf{x}}-\mathbf{x}$ is zero mean cyclic symmetric complex Gaussian noise by numerical validation with two steps, the first step is that the real and imaginary parts of residual error $\hat{\mathbf{x}}-\mathbf{x}$ are both approximate Gaussian distribution and the second step is that the real and imaginary parts on each data-transmitted subcarrier for D/ACO-OFDM is approximately independent. Set $\epsilon_B=1$ and $\epsilon_{top}=2$ for DCO-OFDM, $\epsilon_{top}=2$ for ACO-OFDM and $N=64$, $y_{max}=0.5$, $P_{T_{max}}=0.1$ and $10^5$ symbols for both them. Standard Gaussian kernel density estimation is adopted to obtain the estimated probability density of real and imaginary parts of residual error $\hat{\mathbf{x}}-\mathbf{x}$ with $10^5$ samples. Estimated and moment fitting Gaussian PDF of the real and imaginary parts of residual error $\hat{\mathbf{x}}-\mathbf{x}$ on $1^{th}$ and $31^{th}$ subcarrier for D/ACO-OFDM are shown in Figs. \ref{fig_DCO_1_fit}-\ref{fig_ACO_31_fit}. We can conclude that the PDF of the real and imaginary parts of residual error $\hat{\mathbf{x}}-\mathbf{x}$ are approximate Gaussian with zero mean and of identical distribution. Fig. \ref{fig_Cov_AD} shows the covariance of the real and imaginary parts on each data-transmitted subcarrier for D/ACO-OFDM with less than $10^{-2}$ value. Thus, the real and imaginary parts of residual error $\hat{\mathbf{x}}-\mathbf{x}$ are approximate independent identically Gaussian distribution.
	
	\begin{figure}[]
		\vspace{0cm}
		\setlength{\abovecaptionskip}{0cm} 
		\centering
		{\includegraphics[width =0.9\columnwidth]{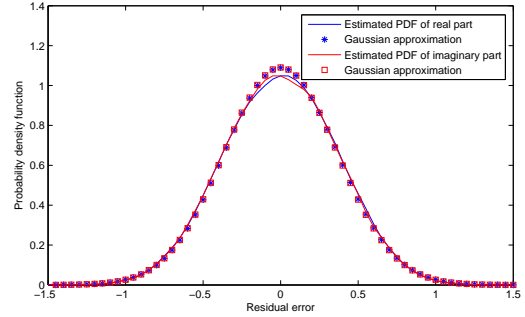}}
		\caption{The Gaussian PDF approximation on the $1^{st}$ subcarrier for DCO-OFDM.}
		\label{fig_DCO_1_fit}
	\end{figure}
	\begin{figure}[]
		\vspace{0cm}
		\setlength{\abovecaptionskip}{0cm} 
		\centering
		{\includegraphics[width =0.9\columnwidth]{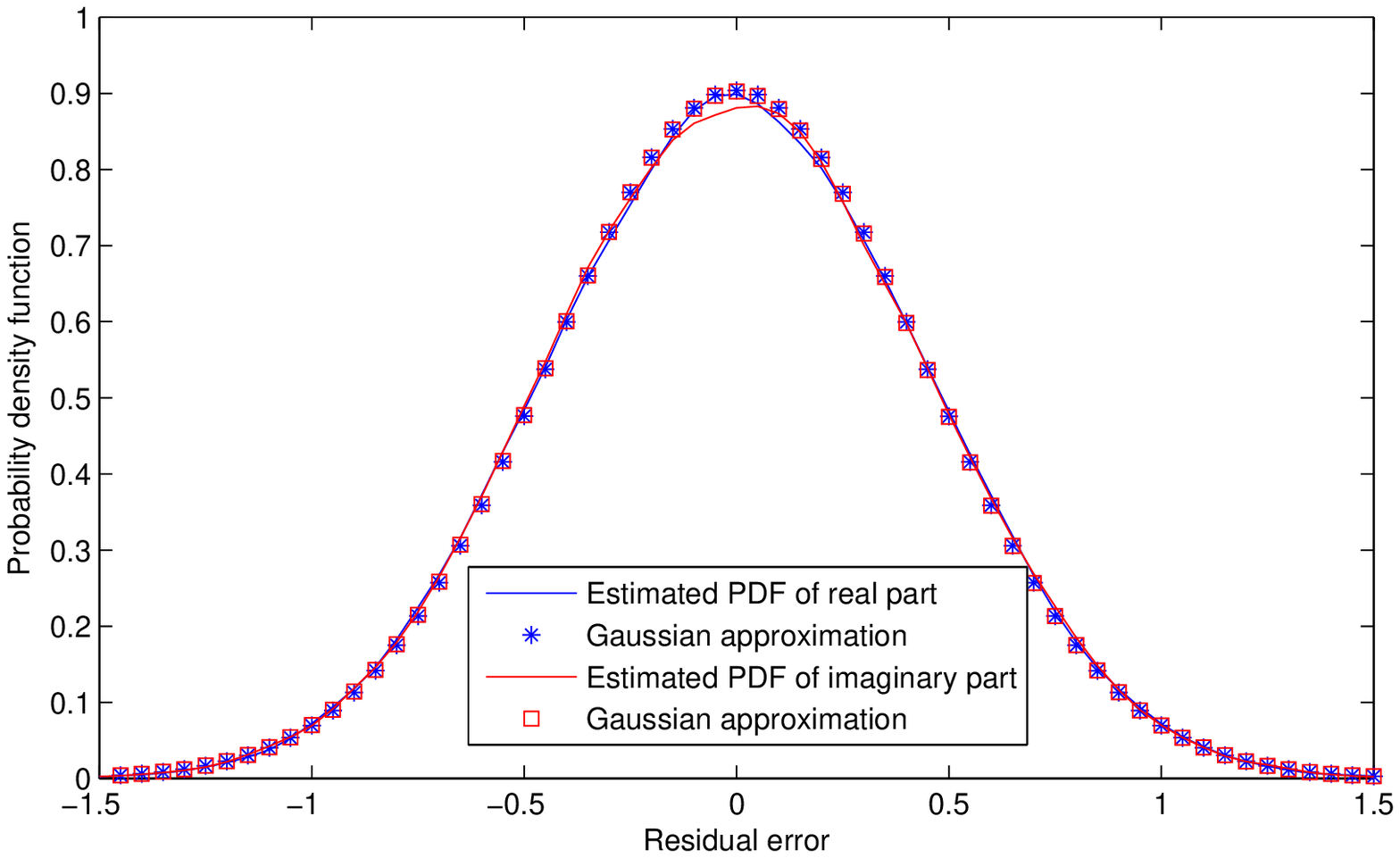}}
		\caption{The Gaussian PDF approximation on the $31^{th}$ subcarrier for DCO-OFDM.}
		\label{fig_DCO_31_fit}
	\end{figure}
	\begin{figure}[]
		\vspace{0cm}
		\setlength{\abovecaptionskip}{0cm} 
		\centering
		{\includegraphics[width =0.9\columnwidth]{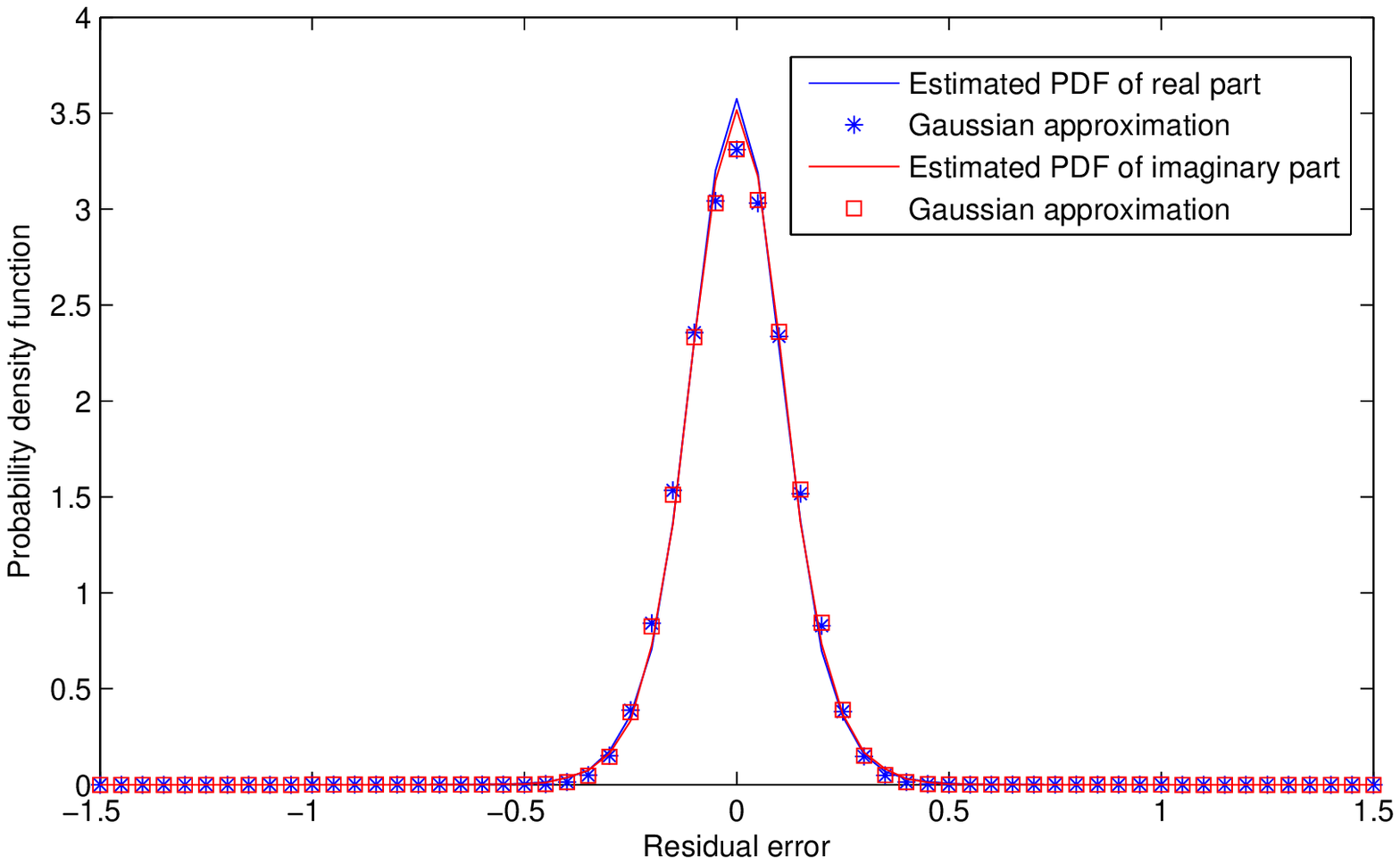}}
		\caption{The Gaussian PDF approximation on the $1^{st}$ subcarrier for ACO-OFDM.}
		\label{fig_ACO_1_fit}
	\end{figure}
	\begin{figure}[]
		\vspace{0cm}
		\setlength{\abovecaptionskip}{0cm} 
		\centering
		{\includegraphics[width =0.9\columnwidth]{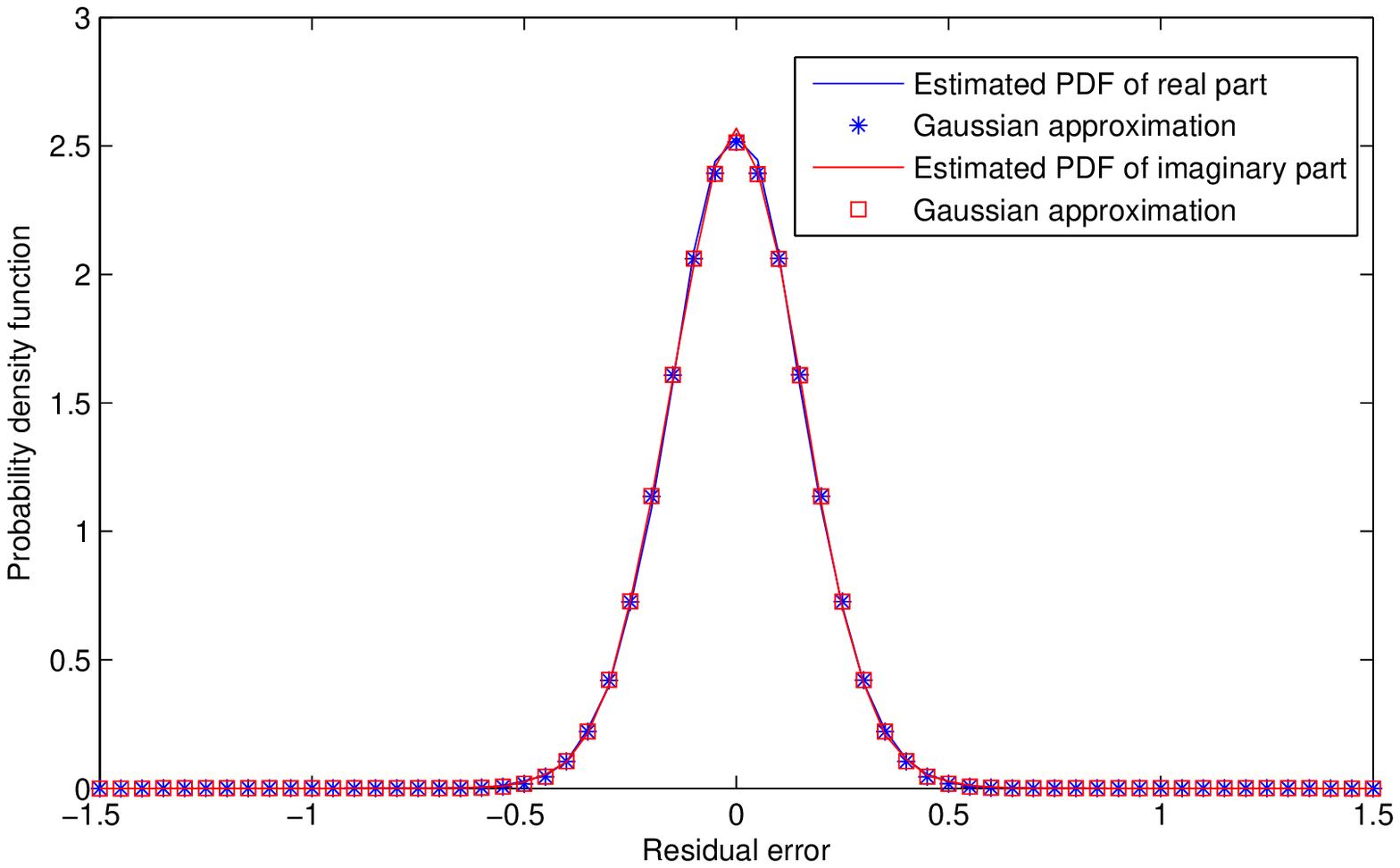}}
		\caption{The Gaussian PDF approximation on the $31^{th}$ subcarrier for ACO-OFDM.}
		\label{fig_ACO_31_fit}
	\end{figure}
	\begin{figure}[]
		\vspace{0cm}
		\setlength{\abovecaptionskip}{0cm} 
		\centering
		{\includegraphics[width =0.9\columnwidth]{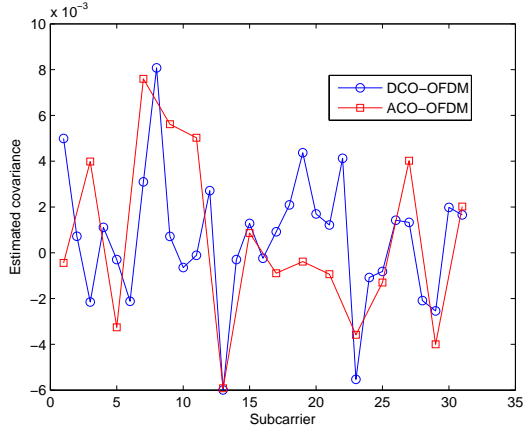}}
		\caption{Estimated covariance of the real and imaginary parts for D/ACO-OFDM  }
		\label{fig_Cov_AD}
	\end{figure}
	
	Furthermore, Table \ref{tab.capa} and Fig. \ref{fig_ApDCOopt} show the optimized total rate obtained from GA and uniform power allocation for DCO-OFDM and ACO-OFDM. The power constraint $P_{Tmax}=0.1$W and peak power varies from $0.05W$ to $1.20W$. We adopt GA due to the nonlinear and nonconvex power allocation problem, and search $\sigma_y$ and $B_{DC}$ for DCO-OFDM and $\sigma_y$ for ACO-OFDM with the same linear range. In GA, we adopt the Matlab GA toolbox designed by University of Sheffield, with parameters in Table \ref{tab.GApar}.
	The next generation samples are selected by stochastic universal selection with different probabilties according to their objective function value. In addition, discrete recombination and real-value mutation is conducted according to Breeder Genetic Algorithm. It is seen that uniform power allocation can perform almostly the same as the power allocation from GA, which can be justified by the high SNR on each subcarrier through optimizing $\epsilon_{B}$ and $\epsilon_{top}$ and a little gain difference, as well as the Poisson noise that increases with the signal power and makes the subcarrier SNR closer to each other. We have also performed power allocation for DCO-OFDM and ACO-OFDM with $128$ subcarriers. It is observed that the total rate for the uniform power allocation is also quite close to that for optimized power allocation from GA. The results are not presented in this four-page letter due to the page limit.
	
	It is also observed from Table \ref{tab.capa} that the peak power for total rate saturation is about $0.25W$ and $1W$ for DCO-OFDM and ACO-OFDM, respectively, where uniform power allocation shows negligible total rate loss in the magnitude of $10^{-2}$ to $10^{-3}$. Larger saturation power for ACO-OFDM can be justified by larger dynamic range of ACO-OFDM compared with DCO-OFDM given the same transmission power, which require larger peak power to guarantee no clipping.  
	\begin{table}[]
		\setlength{\abovecaptionskip}{-0cm} 
		\setlength{\belowcaptionskip}{-0cm}
		\centering
		\caption{The gains of first $32$ subcarriers/$10^{-8}$ }\label{tab.subgain}
		\begin{tabular}{|c|c|c|c|}
			\hline
			1.357+0.000i &
			1.353-0.047i &
			1.341-0.093i &
			1.323-0.135i \\ \hline
			1.298-0.173i &
			1.269-0.205i &
			1.237-0.231i &
			1.203-0.252i \\ \hline
			1.168-0.267i &
			1.133-0.277i &
			1.099-0.282i &
			1.067-0.283i \\ \hline
			1.036-0.280i &
			1.008-0.275i &
			0.981-0.267i &
			0.957-0.257i \\ \hline
			0.935-0.246i &
			0.915-0.234i &
			0.897-0.220i &
			0.881-0.206i \\ \hline
			0.866-0.191i &
			0.853-0.176i &
			0.842-0.161i &
			0.832-0.145i \\ \hline
			0.823-0.129i &
			0.815-0.113i &
			0.809-0.097i &
			0.804-0.081i \\ \hline
			0.799-0.065i &
			0.796-0.049i &
			0.794-0.032i &
			0.792-0.016i \\ \hline
		\end{tabular}
	\end{table}
	\begin{table}[]
		\setlength{\abovecaptionskip}{-0cm} 
		\setlength{\belowcaptionskip}{-0cm}
		\centering
		\caption{The total rate of DCO-OFDM and ACO-OFDM }\label{tab.capa}
		\begin{tabular}{|c|c|c|c|c|}
			\hline
			Peak power & DCO-GA  & DCO-Uniform & ACO-GA  & ACO-Uniform \\ \hline
			0.05W      & 79.519  & 79.439      & 62.686  & 62.686     \\ \hline
			0.10W      & 94.108  & 94.108      & 72.646  & 72.644     \\ \hline
			0.15W      & 103.087 & 103.075     & 78.528  & 78.528     \\ \hline
			0.20W      & 109.637 & 109.624     & 82.728  & 82.724     \\ \hline
			0.25W      & 113.416 & 112.760     & 85.999  & 85.997     \\ \hline
			0.30W      & 113.718 & 113.129     & 88.679  & 88.677     \\ \hline
			0.40W      & 113.716 & 112.640     & 92.921  & 92.921     \\ \hline
			0.50W      & 113.717 & 112.997     & 96.223  & 96.221     \\ \hline
			0.60W      & 113.723 & 113.131     & 98.926  & 98.926     \\ \hline
			0.70W      & 113.723 & 113.417     & 101.216 & 101.216    \\ \hline
			0.80W      & 113.688 & 112.640     & 103.203 & 103.202    \\ \hline
			0.90W      & 113.722 & 113.417     & 104.613 & 104.539    \\ \hline
			1.00W      & 113.716 & 112.997     & 104.958 & 104.845    \\ \hline
			1.10W      & 113.723 & 113.416     & 105.017 & 104.898    \\ \hline
			1.20W      & 113.721 & 113.131     & 105.029 & 104.905    \\ \hline
		\end{tabular}
	\end{table}
	\begin{figure}[]
			\setlength{\abovecaptionskip}{-0.3cm} 
			\setlength{\belowcaptionskip}{-1cm}
			\centering
			{\includegraphics[width =1\columnwidth]{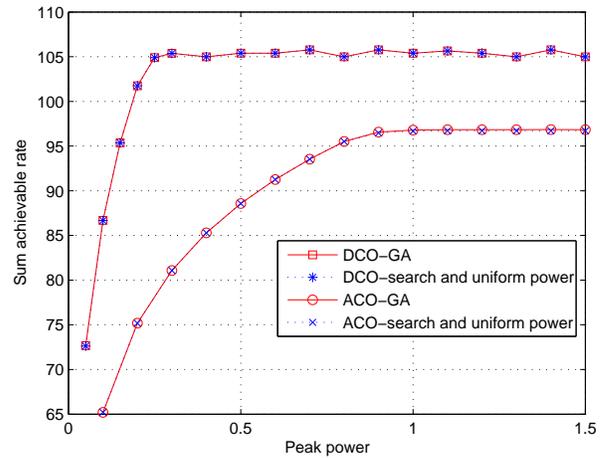}}
			\caption{The sum achievable rate with respect to the peak power.}
			\label{fig_ApDCOopt}
	\end{figure}
 \begin{table}[htbp]
 	\centering
 	\caption{Parameters of GA for DCO-OFDM}
 	\begin{tabular}{c c}
 		\hline  \hline
 		Parameters   &   Value   \\  
 		\hline
 		Number of individuals             &1000 \\
 		Maximum number of generations     &70\\
 		Precision of variables            &20bits\\
 		Generation gap                    &0.9\\
 		Lower bound on the $[w_1,\cdots,w_{31},B_{DC}]$         &0   \\
 		Upper bound on the $[w_1,\cdots,w_{31},B_{DC}]$          &[$0.5$*ones(1,32)]  \\
 		Peak power $y_{max}$                &0.5W\\
 		\hline \hline
 	\end{tabular}
 \end{table}
	\section{Conclusions}\label{sec.Conclusions}
 We have investigated the characteristic of clipping noise for DCO-OFDM and ACO-OFDM in Poisson channel. There exists a balance in terms of $\epsilon_{B}$ and $\epsilon_{top}$ between the clipping noise and Poisson noise, where smaller $\epsilon_{B}$ and $\epsilon_{top}$ may increase the clipping noise and larger $\epsilon_{B}$ and $\epsilon_{top}$ may increase the Poisson noise. Moreover, we have formulated the subcarrier power allocation to maximize the total rate. It is observed that uniform power allocation can achieve virtually the same total rate as the optimized power allocation obtained from GA with significantly reduced computational complexity.
 
 \begin{appendices} 
 \section{Proof of theorem \ref{the.clip}}\label{app.1}
	\begin{proof}
		For DCO-OFDM, the time domain signal $y_n$ can be approximated by Gaussian distribution $\cal N$$(0,\sigma_{y}^{2})$, defined as $f_{y_n}(\cdot)$. Thus, we have the scaling factor 
		\be
		K&=&\frac{\mathbb{E}[y_n^{bias},\hat{y}_n]}{\mathbb{\mathbb{E}}[(y_n^{bias})^2]}\nonumber \\
		&=&\frac{\int_{-\infty}^{+\infty}(x+B_{DC})C(x+B_{DC})f_{y_n}(x)\mathrm{d}x}{\int_{-\infty}^{+\infty}(x+B_{DC})^2f_{y_n}(x)\mathrm{d}x}\nonumber \\
		&=& \{ \epsilon_B[\phi(\epsilon_B)-\phi(\epsilon_{top}-\epsilon_B)]+(1+\epsilon_B^2)Q(-\epsilon_B) \nonumber \\ &&-(1+\epsilon_B^2-\epsilon_{top}\epsilon_B)Q(\epsilon_{top}-\epsilon_B)]\}/{(1+\epsilon_B^2)}.
		\ee
		where $\phi (u)=\frac{1}{\sqrt{2\pi }}\exp(-\frac{u^2}{2}), Q(u)=\int_{u}^{+\infty }\phi (t)\mathrm{d}t$. Define $y_n^{clip}\overset{\Delta}{=}\hat{y}_{n}-y_n^{bias}$.
		According to $\hat{y}_{n}=K\cdot y_n^{bias}+n_c(n)$, we have the following on the clipping noise and its second order moment,
		\be
		n_c(n)&=&(1-K)y_n^{bias}+y_{n}^{clip},\\
		\mathbb{E}[n_c^2(n)]&=&(1-2K+K^2)\mathbb{E}[(y_n^{bias})^2]+\mathbb{E}[(y_n^{clip})^2] \nonumber\\
		&&+2(1-K)\mathbb{E}[y_n^{bias}(\hat{y}_n-y_n^{bias})]\nonumber \\
		&=&(1-K^2)\mathbb{E}[(y_n^{bias})^2]-\mathbb{E}[(y_n^{bias})^2]+\mathbb{E}[(\hat{y}_n)^2]\nonumber  \nonumber\\
		&=& \sigma_y^2[\epsilon_B\phi (-\epsilon_B)-(\epsilon_B+\epsilon_{top})\phi(\epsilon_{top}-\epsilon_B) \nonumber\\
		&&+(1+\epsilon_B^2)Q(-\epsilon_B)+(\epsilon_{top}^2-\epsilon_B^2-1)\cdot  \nonumber\\
		&&Q(\epsilon_{top}-\epsilon_B)]-K^2(\sigma_y^2+B_{DC}^2). 
		\ee	
		
		We analyze the impact of clipping noise $n_c(n)$ to each subcarrier based on the identically and independently distributed assumption. Note that the expectation and variance of $n_c(n)$ are independent on index $n$, we let that $\mu\dff\mathbb{E}[n_c(n)]$ and $\sigma^2\dff\mathbb{D}[n_c(n)]$. Let $n_k$ denote the frequency domain of clipping noise on subcarrier $k$, given as follow
		\be
		n_k&=&\frac{1}{N} \sum_{n=0}^{N-1}n_c(n)e^{-j\frac{2\pi nk}{N}}.\\
		\mu&\dff&\mathbb{E}[n_c(n)]=(1-K)\sigma_y\epsilon_B\nonumber\\&&+\int_{-\infty}^{+\infty}\Big(C(x+B_{DC})-x-B_{DC}\Big)f_{y_n}(x)\mathrm{d}x\nonumber\\
		&=&\sigma_y[(1-K)\epsilon_B(1-\epsilon_B)Q(\epsilon_B)\nonumber \\
		&&+(\epsilon_{top}-\epsilon_{B}-1)Q(\epsilon_{top}-\epsilon_{B})],\\
		\sigma^2&\dff&\mathbb{D}[n_c(n)]=\mathbb{E}[n_c^2(n)]-\mathbb{E}^2[n_c(n)]\nonumber\\&=&\sigma_y^2[\epsilon_B\phi (-\epsilon_B)-(\epsilon_B+\epsilon_{top})\phi(\epsilon_{top}-\epsilon_B) \nonumber\\
		&&+(1+\epsilon_B^2)Q(-\epsilon_B)+(\epsilon_{top}^2-\epsilon_B^2-1)Q(\epsilon_{top}-\epsilon_B)] \nonumber\\
		&&-K^2(\sigma_y^2+B_{DC}^2)-\mu^2,
		\ee  
		We have
		$\mathbb{E}[n_k] = \mu$ for $k = 0$ and $0$ for $k \neq 0$,  and its variance $\mathbb{D}[n_k]=\frac{\sigma^2}{N}.$ Define $y_n^r$ as output signal that clipping signal goes through the linear time invariant system with $k$th subcarrier gains $g_k$. Note that the frequency signal of clipping signal is equivalent to $Kx_k+KB_{DC}\mathbbm{1}\{k=0\}+n_k$, then we have
		\begin{flalign}
		y_n^r &= K\sum_{k=1}^{N-1}g_k x_k e^{j\frac{2\pi nk}{N}}+Kg_0 B_{DC}+\sum_{k=0}^{N-1}g_k n_k e^{j\frac{2\pi nk}{N}}\nonumber \\
		&\dff \tilde{y}_n+\nu_n^r,
		\end{flalign}
		where $\tilde y_n$ denotes the summation of the first two signal terms and $ \nu^r_n$ denotes the third noise term.
		
		Note that received photons number $z_n$ follows the Poisson distribution with parameter $\lambda_n=\alpha y_n^r+\lambda_b$ and The received signal $\mathbb{P}(z_n=\nu)=e^{-\lambda_n}\frac{\lambda_n^{\nu}}{\nu!}$, via basic calculation we have 
		\begin{flalign}
		\mathbb{E}[z_n^2]&=\mathbb{E}[(\alpha y_n^r+\lambda_b)^2+\alpha y_n^r+\lambda_b], \\
		\mathbb{E}[z_n z_m]&=\mathbb{E}[(\alpha y_n^r+\lambda_b)(\alpha y_m^r+\lambda_b)], \text{for}\ m\neq n\\
		\mathbb{E}[\nu_n^r]&=\sum_{k=0}^{N-1}g_k \mathbb{E}[n_k]e^{j\frac{2\pi nk}{N}}=\mu g_0,
		\end{flalign}
		\begin{flalign}
		\mathbb{E}[\nu_n^r(\nu_m^r)^*]&=\mathbb{E}[\sum_{k=0}^{N-1}g_k n_k e^{j\frac{2\pi nk}{N}}\sum_{k^{'}=0}^{N-1}g_{k^{'}}^* n_{k^{'}}^* e^{-j\frac{2\pi n k^{'}}{N}}]\nonumber \\&=\frac{\sigma^2}{N}\sum_{k=0}^{N-1}|g_k|^2e^{j\frac{2\pi k(n-m)}{N}}+|g_0|^2\mu^2.
		\end{flalign}
		Moreover, we have the following on $y^r_n$ and $z_n$,
		\begin{flalign}\label{eq.ynrvar}
		\mathbb{D}[y_n^r]&=\mathbb{E}[|\tilde{y_n}+\nu_n^r|^2]-|\mathbb{E}[\tilde{y_n}+\nu_n^r]|^2\nonumber \\
		&=\mathbb{E}[|\nu_n^r|^2]-\mu^2g_0^2=\frac{\sigma^2}{N}\sum_{k=0}^{N-1}|g_k|^2,\\
		\mathbb{D}[z_n]&=\alpha^2(\mathbb{E}[|y_n^r|^2]-|\mathbb{E}[y_n^r]|^2)+\alpha \mathbb{E}[y_n^r]+\lambda_b \nonumber \\
		&=\alpha^2\frac{\sigma^2}{N}\sum_{k=0}^{N-1}|g_k|^2+\alpha g_0(KB_{DC}+\mu)\nonumber \\
		&\quad+\alpha K\sum_{k=0}^{N-1}g_k x_k e^{j\frac{2\pi kn}{N}}+\lambda_b,
		\end{flalign}		
		\begin{flalign}
		\mathbb{E}[z_n z_m]-\mathbb{E}[z_n]\mathbb{E}[z_m]
		&=\alpha^2(\mathbb{E}[y_n^r (y_m^r)]-\mathbb{E}[y_n^r]\mathbb{E}[y_m^r])\nonumber \\
		&= \alpha^2(\mathbb{E}[\nu_n^r \nu_m^r]-\mathbb{E}[\nu_n^r]\mathbb{E}[\nu_m^r])\nonumber \\
		&= \frac{\sigma^2}{N}\sum_{k=0}^{N-1}|g_k|^2e^{j\frac{2\pi k(n-m)}{N}}.	
		\end{flalign}		
		Thus the variance of $\hat x_k$ on subcarrier $k$ is given by 
		
		\begin{flalign}\label{eq.xnvar}			
		\mathbb{D}[\hat{x}_{k}]&=\frac{1}{N^2}\mathbb{E}[\sum_{n=0}^{N-1}z_n^2+\sum_{n\neq m}z_n z_m e^{-j\frac{2\pi k(n-m)}{N}}]\nonumber \\&\quad-\frac{1}{N^2}(\sum_{n=0}^{N-1}\mathbb{E}^2[z_n]+\sum_{n\neq m}\mathbb{E}[z_n]\mathbb{E}[z_m]e^{-j\frac{2\pi k(n-m)}{N}})\nonumber \\
		&=\frac{1}{N^2}\sum_{n=0}^{N-1}[\alpha g_0(KB_{DC}+\mu)+\alpha K\sum_{k=0}^{N-1}g_k x_k e^{j\frac{2\pi kn}{N}}+\lambda_b]\nonumber\\&\quad+\frac{1}{N^2}\sum_{n=0}^{N-1}\sum_{m=0}^{N-1}\frac{\alpha^2\sigma^2}{N}\sum_{k^{'}=0}^{N-1}|g_{k^{'}}|^2e^{j\frac{2\pi (n-m)(k^{'}-k)}{N}}\nonumber \\
		&=\frac{1}{N}[\alpha g_0(KB_{DC}+\mu)+\lambda_b]+\frac{\alpha^2\sigma^2}{N}|g_k|^2.
		\end{flalign}
		
		For ACO-OFDM, the derivations are similar to that for DCO-OFDM. Note that time domain signal $y_n$ is odd symmetric for ACO-OFDM, bottom clipping signal $y_n^{b}\dff y_n\mathbbm{1}\{y_n\geq0\}=\frac{1}{2}y_n+d_n$, where $d_n=\frac{1}{2}|y_n|$. According to the CLT, the non-distorted time domain signal follows approximated Gaussian distribution for large N and then bottom clipping signal close to the truncated Gaussian distribution $f_{y_n^b}(x)=\frac{1}{\sqrt{2\pi\sigma_y^2}}e^{-\frac{x^2}{2\sigma_y^2}}U(x)+\frac{1}{2}\delta(x)$, where $U(x)$ and $\delta(x)$ are step function and Dirac function. Considering top clipping, double-side clipping signal $\hat{y}_n=C(y_n^{b})=Ky_n^{b}+n_c(n)$ according to Bussgang theorem, where 
		\be 
		K&=&\frac{\mathbb{E}[y_n^{b},\hat{y}_n]}{\mathbb{\mathbb{E}}[(y_n^{b})^2]}
		=\frac{\int_{-\infty}^{+\infty}xC(x)f_{y_n^{b}}(x)\mathrm{d}x}{\int_{-\infty}^{+\infty}x^2f_{y_n^b}(x)\mathrm{d}x}\nonumber \\
		&=&1-2Q(\epsilon_{top}),
		\ee 
		Define $y_n^{clip}=\hat{y}_n-y_n^b$, $\mu\dff\mathbb{E}[n_c(n)]$ and $\sigma^2=\mathbb{D}[n_c(n)]$. Similarly, we have
		\be 
		\mu&=&(1-K)\mathbb{E}[y_n^b]+\mathbb{E}[y_n^{clip}]\nonumber\\
		&=&\frac{(1-K)\sigma_y}{\sqrt{2\pi}}+\int_{-\infty}^{+\infty}\Big(C(x)-x\Big)f_{y_n^b}(x)\mathrm{d}x\nonumber\\
		&=&\sigma_y[-\phi (\epsilon_{top})+\epsilon_{top}Q(\epsilon_{top})+\frac{1-K}{\sqrt{2\pi}}];
		\ee
		\be
		\sigma^2&=&\mathbb{E}[n_c^2(n)]-\mu^2\nonumber\\
		&=&(1-K^2)\mathbb{E}[(y_n^{b})^2]-\mathbb{E}[(y_n^{b})^2]+\mathbb{E}[(\hat{y}_n)^2]-\mu^2  \\
		&=&\sigma_y^{2}[\frac{1-K^2}{2}+(\epsilon_{top}^2-1)Q(\epsilon_{top})-\epsilon_{top}\phi(\epsilon_{top})]-\mu^2.\nonumber
		\ee 
		Thus, we have $\hat{y}_n=\frac{K}{2}y_n+Kd_n+n_c(n)$. Define $D_k$ and $n_k$ as FFT of $d_n$ and $n_c(n)$, respectively. 
		We have the following on the clipping,
		\begin{flalign}
		y_n^r&=\frac{K}{2}\sum_{k=0}^{N-1}g_k x_k e^{j\frac{2\pi nk}{N}}+K\sum_{k=0}^{N-1}g_k D_k e^{j\frac{2\pi nk}{N}}\nonumber \\
		&\quad+\sum_{k=0}^{N-1}g_k n_k e^{j\frac{2\pi nk}{N}}.
		\end{flalign}
		Moreover, we have the following
		\be
		\mathbb{D}[z_n]=\alpha^2\frac{\sigma^2}{N}\sum_{k=0}^{N-1}|g_k|^2+\frac{\alpha K}{2}\sum_{k=0}^{N-1}g_k x_k e^{j\frac{2\pi kn}{N}}\quad\quad\nonumber \\
		+K\sum_{k=0}^{N-1}g_k D_k e^{j\frac{2\pi kn}{N}}+\alpha g_0\mu+\lambda_b,\quad\quad\\
		\mathbb{E}[z_n z_m]-\mathbb{E}[z_n]\mathbb{E}[z_m]=\frac{\sigma^2}{N}\sum_{k=0}^{N-1}|g_k|^2e^{j\frac{2\pi k(n-m)}{N}}.
		\ee
		Note that $y_n=-y_{n+\frac{N}{2}}$ for $0<n<\frac{N}{2}$, we have
		\be 
		 D_k&=&\frac{1}{N}\sum_{n=0}^{N-1}d_n e^{j\frac{2\pi nk}{N}}\nonumber\\&=&\sum_{n=0}^{\frac{N}{2}-1}\frac{|y_n|}{2N} \big(e^{j\frac{2\pi nk}{N}}+e^{j\frac{2\pi(n+\frac{N}{2})k}{N}}\big)=0, \text{ for odd } k.\nonumber
		\ee
		Similar to Equation (\ref{eq.xnvar}), we have
		\be
		\mathbb{D}[\hat{x}_{k}]=\frac{1}{N}\Big(\alpha^2\sigma^2|g_k|^2+\alpha g_0(K\frac{\sigma_y}{\sqrt{2\pi}}+\mu)+\lambda_b\Big).
		\ee
	\end{proof} 
 	\section{Proof of lemma \ref{lemma.noncon}}\label{lemB}
 	\begin{proof}
 		Firstly, it is shown that the constraints of optimization problems for DCO-OFDM is non-convex and function $P_{DCO}(\epsilon_B)\dff B_{DC}+\beta_{DCO}$ is non-convex with respect to $\epsilon_B$ given $\bf w$ for any $\bf{w}$$=(w_1,\cdots,w_{N/2-1})$ as follows.
 		
		Noting that $Q^{'}(x)=-\phi(x)$, $\phi^{'}(x)=-x\phi(x)$ and $\phi^{''}(x)=(x^2-1)\phi(x)$, we have
		\be 
		P_{DCO}^{'}(\epsilon_B)&=&\sigma_y\{1+\phi^{'}(\epsilon_B)+\phi^{'}(\epsilon_{top}-\epsilon_B)\nonumber\\&&-[(\epsilon_{top}-\epsilon_B)\phi^{'}(\epsilon_{top}-\epsilon_B)+\phi(\epsilon_{top}-\epsilon_B)]\nonumber\\&&-Q(\epsilon_B)+\epsilon_B\phi(\epsilon_B)\},
		\ee 
		\be 
		P_{DCO}^{''}(\epsilon_B)&=&\sigma_y\{\phi^{''}(\epsilon_B)-\phi^{''}(\epsilon_{top}-\epsilon_B)\nonumber\\&&+(\epsilon_{top}-\epsilon_B)\phi^{''}(\epsilon_{top}-\epsilon_B)\nonumber\\&&+2\phi^{'}(\epsilon_{top}-\epsilon_B)+2\phi(\epsilon_B)+\epsilon_B\phi^{'}(\epsilon_B)\}\nonumber\\
		&=&\sigma_y\{\phi(\epsilon_B)+(\epsilon_{top}-\epsilon_B-1)\phi^{''}(\epsilon_{top}-\epsilon_B)\nonumber\\&&+2\phi^{'}(\epsilon_{top}-\epsilon_B)\}.
		\ee 
		Set $\epsilon_B=\epsilon_{top}-1$, we have $P_{DCO}^{''}(\epsilon_{top}-1)=\sigma_y\{\phi(\epsilon_{top}-1)-2\phi(1)\}<\sigma_y\{\phi(0)-2\phi(1)\}<0$. Thus, function $P_{DCO}(\epsilon_B)$ is non-convex with respect to $\epsilon_B$ given $\bf w$, i.e., the constraint function is non-convex. 
 		
 		Moreover, the constraint of optimization problems for ACO-OFDM is non-convex. Function $P_{ACO}({\bf w})\dff \frac{\sigma_y}{\sqrt{2\pi}}+\beta_{ACO}$ is non-convex with respect to $\bf w$.
 		
		Noting that $\sigma_y\epsilon_{top}=y_{max}$ (constant for system), we have
		\be 
		\frac{\partial P_{ACO}}{\partial \sigma_y}&=&\frac{1}{\sqrt{2\pi}}+\frac{y_{max}^2}{\sigma_y^2}\phi(\frac{y_{max}}{\sigma_y})-\phi(\frac{y_{max}}{\sigma_y})\nonumber\\&&+\frac{y_{max}}{\sigma_y}\phi^{'}(\frac{y_{max}}{\sigma_y})\nonumber\\
		&=&\frac{1}{\sqrt{2\pi}}-\phi(\frac{y_{max}}{\sigma_y}),\\
		\frac{\partial^2 P_{ACO}}{\partial \sigma_y^2}&=&\frac{y_{max}}{\sigma_y^2}\phi^{'}(\frac{y_{max}}{\sigma_y})<0.
		\ee 
		Note that for composite function $f(g(x))$, we have $\frac{\partial^2 f(g(x))}{\partial x^2}=f^{''}(g(x))g^{'}(x)^2+f^{'}(g(x))g^{''}(x)$. Since $\sigma_y=\sqrt{\sum w_i^2}$ and $\frac{\partial^2 \sigma_y}{\partial w_j^2}=\frac{\sum_{i\neq j} w_i^2}{(\sum w_i^2)^{\frac{3}{2}}}$, setting $w_i=0$ for $i\neq j$, we have $\frac{\partial^2 P_{ACO}({\bf w})}{\partial w_j^2}=\frac{\partial^2 P_{ACO}}{\partial \sigma_y^2}(\frac{\partial \sigma_y}{\partial w_j})^2<0$. Thus, function $P_{ACO}({\bf w})$ is non-convex with respect to $\bf w$, i.e., the constraint function is non-convex. 

	\end{proof}
\end{appendices}

	\bibliographystyle{IEEEtran}
	\bibliography{OFDM2}

\end{document}